\documentclass[conference]{IEEEtran}
\IEEEoverridecommandlockouts

\usepackage{cite}
\usepackage{amsmath,amssymb,amsfonts}
\usepackage{algorithm}
\usepackage{algorithmicx}
\usepackage{algpseudocode}
\usepackage{graphicx}
\usepackage{textcomp}
\usepackage{xcolor}
\usepackage{url}
\usepackage{float}
\usepackage{booktabs}
\usepackage{graphicx}
\usepackage{subcaption}
\usepackage{xurl}
\usepackage{hyperref}
\usepackage{fancyhdr}
\def\BibTeX{{\rm B\kern-.05em{\sc i\kern-.025em b}\kern-.08em
    T\kern-.1667em\lower.7ex\hbox{E}\kern-.125emX}}
\usepackage{pifont}
\newcommand{\circnum}[1]{\ding{\the\numexpr171+#1\relax}}

\usepackage{xcolor}
\definecolor{darkgreen}{HTML}{006400}
\newcommand{\new}[1]{#1}

\newcommand{\edit}[1]{#1}

\newcommand{\reviewed}[1]{{#1}}
\makeatletter
\newcommand{\linebreakand}{%
  \end{@IEEEauthorhalign}
  \hfill\mbox{}\par
  \mbox{}\hfill\begin{@IEEEauthorhalign}
}
\makeatother

\begin{document}

\title{Cheaper by the Batch: Shared Traversal for Genotype Graph Editing}

\author{\IEEEauthorblockN{Aaron Li}
\IEEEauthorblockA{
\textit{Cornell University}\\
Ithaca, NY, USA \\
al2633@cornell.edu}
\and
\IEEEauthorblockN{Yifan Li}
\IEEEauthorblockA{
\textit{Cornell University}\\
Ithaca, NY, USA \\
yl3722@cornell.edu}
\and
\IEEEauthorblockN{Drew DeHaas}
\IEEEauthorblockA{
\textit{Cornell University}\\
Ithaca, NY, USA \\
dcd239@cornell.edu}
\and
\IEEEauthorblockN{Giulia Guidi}
\IEEEauthorblockA{
\textit{Cornell University}\\
Ithaca, NY, USA \\
gg434@cornell.edu}
}

\definecolor{authorGiulia}{HTML}{C71585}
\definecolor{authorYifan}{HTML}{8B008B}
\newcommand{\giulia}[1]{\textcolor{authorGiulia}{Giulia: #1}}
\newcommand{\yifan}[1]{\textcolor{authorYifan}{Yifan: #1}}

\maketitle
\thispagestyle{fancy}
\lhead{}
\rhead{}
\chead{}
\lfoot{\footnotesize{
SC26 Workshops, November 15-20, 2026, Chicago, Illinois, USA
\newline 979-8-3195-1221-5/26/\$31.00 \copyright 2026 IEEE}}
\rfoot{}
\cfoot{}
\renewcommand{\headrulewidth}{0pt}
\renewcommand{\footrulewidth}{0pt}
\begin{abstract}
Updating a graph by inserting or replacing nodes while preserving semantics and reusing existing structure is a recurring computational problem. 
In population genetics, this problem arises in the genotype representation graph (GRG), a directed acyclic graph that losslessly encodes phased genetic variation across hundreds of thousands of samples by sharing subgraph structure for individual mutations.
In a GRG, each mutation’s carrier set is implicitly encoded as the set of leaf nodes reachable from the node it is assigned to. 
Updating a mutation is therefore a structural editing problem, and current approaches remap mutations individually.

This paper introduces a batched mutation-remapping algorithm that replaces independent reuse-aware traversals with a single shared reverse-topological pass, identifying reuse candidates for an entire batch at once.
The pass propagates compact bit-parallel per-mutation state and uses an adaptive sparse/dense carrier set representation spanning rare-to-common variant densities.
Batching is the memory-scalable complement to split-based parallelism, which instead replicates graph and traversal state per worker. 
Our remapping is evaluated on a controlled update workload and on end-to-end allele polarization, a bulk carrier set update that is common in population genetic analysis. 
Our approach is up to 10.5$\times$ faster than independent remapping while preserving exact carrier-set semantics.
\end{abstract}


\section{Introduction}
\label{sec:introduction}

Biobank datasets contain the genomes of hundreds of thousands of people, each differing from the reference at millions of sites~\cite{all2019all, bycroft2018uk}.
This variation is often stored in compressed dense matrix formats such as VCF~\cite{danecek2011variant_vcf} or BCF~\cite{li2011statistical_bcf}.
As sample sizes grow, these representations scale poorly because they cannot capture the extensive structure shared across genetically similar samples. 
Recent formats such as PLINK PGEN store data sparsely and leverage the commonality between samples to compress it in place~\cite{plink}. 
This commonality, however, is generated by a hierarchical process, namely the genealogy of the individuals, yet most formats fail to exploit hierarchical sharing for compression.
Recent work addresses this with the genotype representation graph (GRG), a multitree directed acyclic graph that losslessly encodes phased genetic variation across hundreds of thousands of samples by sharing subgraph structure across mutations~\cite{dehaas2024grg,dehaas2026general}.

This shared substructure is what makes GRGs compact, but it also means the information is baked into the graph’s topology. 
A GRG does not store which samples carry a mutation; the carrier set is implied by the leaves reachable from the node to which the mutation is assigned. 
Editing a GRG therefore fundamentally differs from updating a sparse genotype matrix, where the carriers of a mutation are revised by rewriting a column.
Downsampling to a subset of samples or mutations is straightforward, but changing which samples carry a mutation is not a local edit. 
The carrier set is encoded through reachability, so altering it requires restructuring the graph. 
Because post-construction revision is difficult, the typical workflow builds a graph from a fixed callset and treats it as a static substrate for downstream analysis.

Consider a GRG constructed from an initial set of mutations. 
Over time, updates may affect only a subset of them, such as corrections to carrier sets or the addition of newly identified mutations. 
The core operation underlying such updates is \emph{mutation remapping}: given a mutation and its desired carrier set, attach it to reusable graph structure when possible, and introduce new structure only when necessary. This is an irregular graph workload: each mutation induces a different, data-dependent traversal whose size and shape depend on its carrier set. As a result, different mutations visit different parts of the graph and require different amounts of work.

Our proposed batched formulation solves this at scale and enables post-construction GRG editing.
Rather than treating each mutation independently, it processes a batch of edits in a single traversal of the existing graph, propagating compact, mutation-specific state to identify reusable structures for many mutations at once. 
An adaptive sparse/dense carrier-set representation, mirroring the sparse accumulator choice in sparse linear algebra, keeps that state cheap across a range of densities. 
By amortizing traversal and carrier-set processing across a batch, updates become more efficient, and the strategy composes with existing parallel mutation mapping, enabling traversal sharing and parallel execution together.

\new{The representative editing task in the evaluation is allele polarization, a population genetics operation that reorients each variant based on which allele is ancestral. 
Polarization rewrites carrier sets in bulk across many sites, exercising the reuse-aware remapping this method targets.
In Section~\ref{sec:allele-polarization}, we develop the reduction to mutation remapping.}

Our contributions are:
\begin{itemize}
\item A mutation remapping algorithm for editing a genotype representation graph after construction via a single shared traversal over a batch of carrier set updates;
\item A compact bit-parallel representation that reduces per-mutation state overhead during shared traversal, improving memory efficiency;
\item An end-to-end evaluation of polarization, achieving up to 10.5$\times$ speedup while preserving carrier-set semantics.
\end{itemize}

\reviewed{The implementation for mutation mapping is written in C++, while the polarization-specific code is written in Python. Both can be found at} \url{https://github.com/CornellHPC/GRG-shared-traversal}.
\section{Background}
\label{sec:background}

\subsection{Genotype Representation Graphs}
\label{sec:grg-background}

A genotype representation graph (GRG) is a leaf-labeled directed acyclic graph that losslessly represents phased genetic variation across a set of individuals~\cite{dehaas2024grg}.
Each leaf represents a \emph{haplotype}: the sequence of alleles along a single inherited chromosome copy, so a diploid individual contributes two haplotypes.
In this paper, we use \emph{sample}, \emph{haplotype}, and \emph{leaf} interchangeably: each refers to one phased chromosome copy, and each is a leaf of the GRG.

The internal nodes represent groups of haplotypes that share a descendant structure. 
GRGs have a multitree topology: internal nodes can have multiple parents, but there is at most one directed path between any pair of nodes. 
A GRG is not unique for a given dataset; many topologies encode the same carrier sets, creating opportunities for reuse-aware editing.

Intuitively, a GRG is a maximally shared DAG: when many haplotypes carry the same group of mutations, that group is represented by a single internal node reused by all of them, much as a hash-consed data structure stores each distinct subtree once and lets multiple parents point to it~\cite{van2000efficient, filliatre2006type}.

For a graph node $n$, let $D(n)$ denote the set of sample leaves reachable from $n$.
A mutation $m$ with carrier set $S_m$, meaning each sample in $S_m$ carries mutation $m$, is assigned to a node $n_m$ such that:

\begin{equation}
D(n_m) = S_m.
\label{eq:mutation-reachability}
\end{equation}

\noindent
That is, every sample reachable from $n_m$ carries $m$, and every sample carrying $m$ is reachable from $n_m$. Notably, multiple mutations may be assigned to the same node when they share identical carrier sets, while mutations with partially overlapping carrier sets can reuse portions of the same descendant graph structure.

\subsection{Mutation Mapping}
\label{sec:b-mapping}

Given a mutation $m$ with carrier set $S_m$, \texttt{MapMutations} assigns $m$ to a node that satisfies the GRG reachability invariant, reusing existing graph structure where possible to preserve compactness.

The \texttt{MapMutations} procedure processes mutations individually in two stages: it first discovers existing graph nodes that can represent subsets of $S_m$, and then uses those candidates to place each mutation.

\subsubsection{Candidate Discovery}
\label{sec:individual-candidate-discovery}

Candidate discovery performs a reverse topological traversal initialized from the carrier samples.
For each visited node $n$, the mapper determines whether its descendant sample set is contained in the target carrier set.
A node is compatible with $m$ when:
\begin{equation}
D(n) \subseteq S_m.
\label{eq:single-mutation-compatibility}
\end{equation}

\noindent
A compatible node represents a group of target carriers without including any non-carrier samples. 
It can therefore be reused when constructing a node for $m$.
The containment condition also determines when traversal can stop. If a visited node reaches any sample outside $S_m$, that node cannot be reused for $m$.
Moreover, every ancestor of that node reaches the same non-carrier sample, so traversal does not continue along that branch. 
By contrast, a compatible node is recorded as a candidate, and its parents are visited to identify larger potential candidate subgraphs.

Candidate discovery therefore produces existing nodes whose descendant sample sets are subsets of the desired carrier set. 
These candidates vary in size and may overlap or be nested within the graph. 
The application stage then selects a compatible collection covering as much of $S_m$ as possible without including samples outside the target set.

\subsubsection{Mutation Application}
\label{sec:mutation-application}

\new{A candidate whose descendant set exactly matches $S_m$ receives the mutation directly, with no modification to the graph.
Otherwise, candidates are sorted by decreasing $|D(n)|$ and selected greedily whenever they are disjoint from all previously selected candidates. 
A new mutation node is created above the selected set, and any carriers in $S_m$ that are left uncovered attach directly.
This prioritizes reuse of the largest compatible subgraphs while guaranteeing that the node reaches exactly $S_m$.}

\begin{figure}[t]
    \centering

    \begin{subfigure}[t]{0.48\linewidth}
        \centering
        \includegraphics[width=\linewidth]{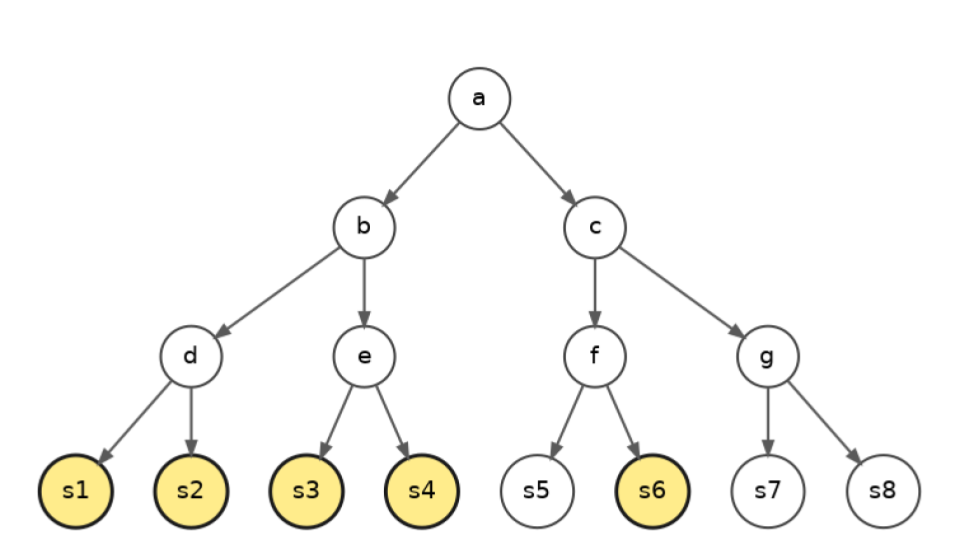}
        \caption{}
        \label{fig:singlemm-p1}
    \end{subfigure}
    \hfill
    \begin{subfigure}[t]{0.48\linewidth}
        \centering
        \includegraphics[width=\linewidth]{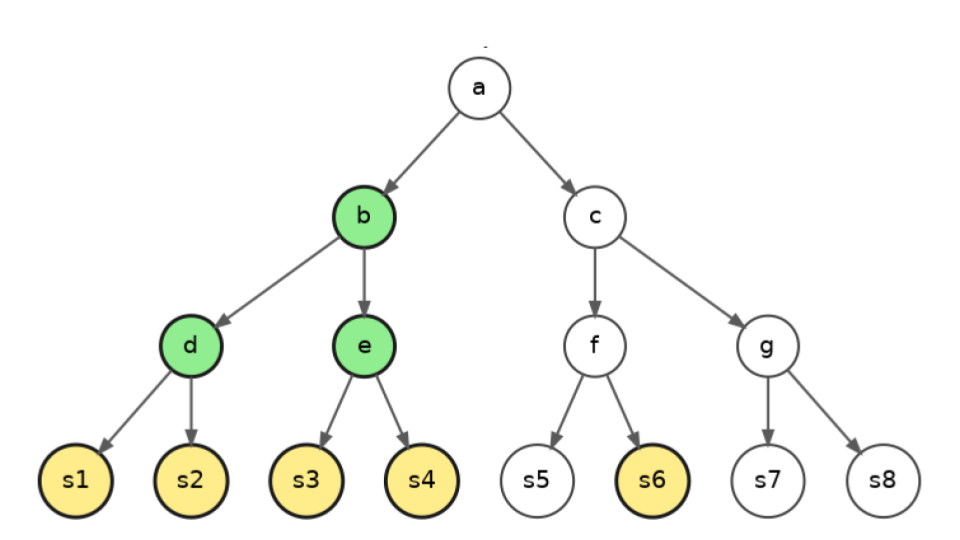}
        \caption{}
        \label{fig:singlemm-p2}
    \end{subfigure}

    \vspace{0.5em}

    \begin{subfigure}[t]{0.48\linewidth}
        \centering
        \includegraphics[width=\linewidth]{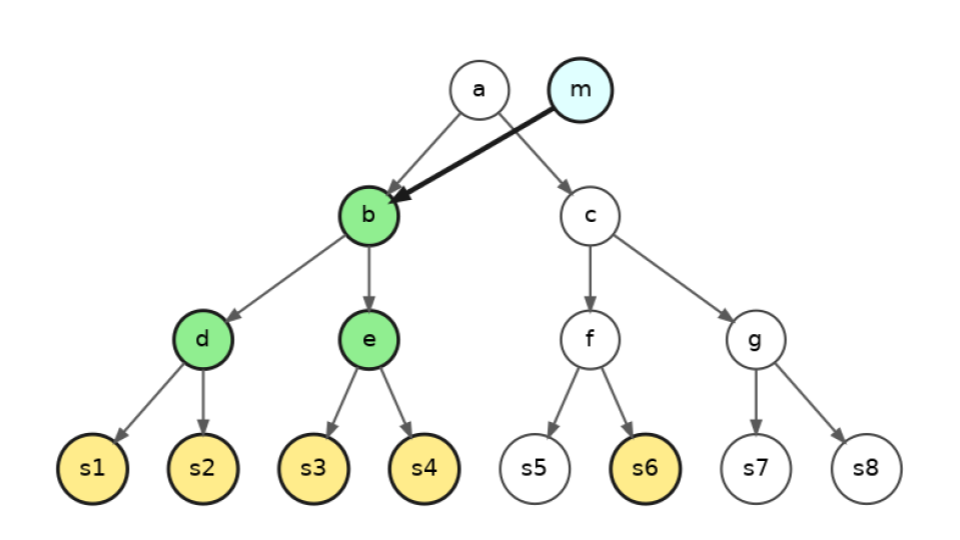}
        \caption{}
        \label{fig:singlemm-p3}
    \end{subfigure}
    \hfill
    \begin{subfigure}[t]{0.48\linewidth}
        \centering
        \includegraphics[width=\linewidth]{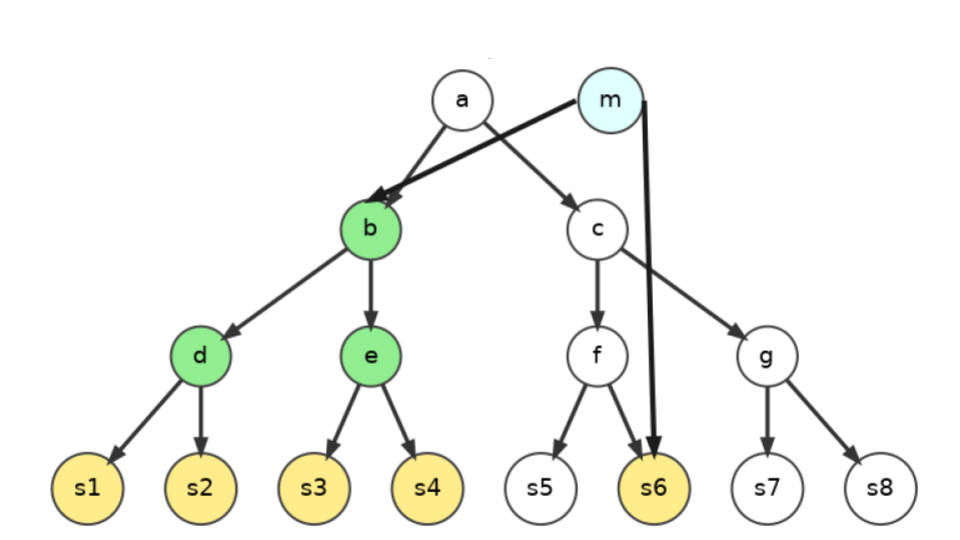}
        \caption{}
        \label{fig:singlemm-p4}
    \end{subfigure}

    \caption{Overview of the mutation mapping procedure. In \protect\subref{fig:singlemm-p1}, traversal is initialized at the carrier samples. In \protect\subref{fig:singlemm-p2}, the graph is traversed upward to identify compatible nodes. In \protect\subref{fig:singlemm-p3}, a new mutation node \(m\) is introduced and connected to the selected candidates. Finally, in \protect\subref{fig:singlemm-p4}, the remaining samples are attached to the new mutation node.}
    \label{fig:singlemm}
\end{figure}

\subsection{Post-Construction Mutation Remapping}
\label{sec:post-construction-remapping}

\new{An update either introduces a new mutation or replaces one whose carrier set has changed. 
In both cases, the graph must be updated so each mutation attaches to a node whose reach is exactly its desired carrier set, reusing existing structure where possible and adding nodes only when the target cannot be represented otherwise. 
A direct approach invokes the single-mutation procedure independently for each update.
\reviewed{Each mutation performs its own candidate-discovery traversal and application step, repeatedly processing shared portions of the same DAG.}
In Section~\ref{sec:methods}, we formalize the batch of updates and eliminate this redundancy.}

\subsection{Allele Polarization}
\label{sec:allele-polarization}

Polarization is a preprocessing operation in population genetic analysis that determines which allele at each variant site is ancestral and which is derived. 
Consistently representing the derived allele enables downstream analyses of population history, including demographic inference and the detection of natural selection~\cite{gutenkunst2009inferring,fay2000hitchhiking}.

A GRG is constructed relative to a reference sequence: alleles that match the reference are represented implicitly, while alternate alleles are represented as mutations.
Polarization replaces this orientation with one based on an ancestral reference sequence, so that the derived allele is represented as the mutation.
If the ancestral allele differs from the current reference allele, the recorded mutation and its carrier set must be updated.

Reorientation is where polarization meets graph editing.
For a biallelic variant, every haplotype takes one of two values, so membership is binary, and the alternate-allele carrier set has a well-defined complement.
In a tabular representation, reorienting is a relabel-and-complement of one binary column; in a GRG, the allele is tied to its carrier set, so the same edit becomes a structural remapping. 
For a biallelic site, let $C$ be the set of haplotypes with an observed (non-missing) allele, and let $S_{\text{alt}} \subseteq C$ be the current alternate carriers.
If the alternate allele is inferred to be ancestral, reorienting swaps the two sets, and the new carrier set is the relative complement within $C$:
\begin{equation}
 S'_{\text{alt}} = C \setminus S_{\text{alt}}.
 \label{eq:complement}
\end{equation}

The original mutation is removed and replaced with one mapped to this complementary set.
Polarization is thus an example of post-construction remapping: flipping a mutation’s ancestral state complements its carrier set, and the replacement must be reattached to a node that reaches exactly the complemented target, reusing existing nodes where such reachable sets already exist and introducing new structure otherwise.
\section{Methods}
\label{sec:methods}

\new{The single-mutation procedure in Section~\ref{sec:b-mapping} solves the $k = 1$ case of remapping. 
The batched remapping problem generalizes this to many updates at once: given a GRG $G$ and updates $U = \{(m_1, S_1), \dots, (m_k, S_k)\}$, produce a GRG in which each $m_i$ attaches to a node that reaches exactly $S_i$, reusing existing structure where possible.
This section focuses on the regime $k \gg 1$.}

\new{Running the single-mutation procedure independently launches $k$ separate traversals over the same GRG, repeatedly reprocessing shared nodes.
This repeated work increases sharply with batch size (quantified in Section~\ref{sec:results}).
Eliminating it motivates replacing the $k$ independent reuse-aware traversals with a single shared traversal processed in two phases: \circnum{1} a shared read-only pass that finds candidate nodes for every mutation in the batch, and \circnum{2} a pass that applies the mutations serially using the size-ordered greedy attachment from Section~\ref{sec:mutation-application}.}

\subsection{Batched Candidate Discovery}
\label{sec:batched-candidate-discovery}

This step finds reuse candidates for every mutation in a single traversal, visiting each active node once instead of once per mutation.
\new{Candidate discovery is initialized from the union of the carrier sets,
\begin{equation}
 S_B = \bigcup_{i=1}^{k} S_i,
\end{equation}
and uses a shared reverse topological traversal from the corresponding leaves.
The traversal relies on the fact that mutation compatibility is inherited from a node’s children.
Given $D(n) = \bigcup_{c \in \text{children}(n)} D(c)$, for any mutation $m_i$,
\begin{equation}
D(n) \subseteq S_i \iff \forall c \in \text{children}(n),\ D(c) \subseteq S_i.
\label{eq:recurrence}
\end{equation}}

\vspace{-.75em}
A node is compatible with $m_i$ if and only if all of its children are.
This recurrence holds independently for each mutation, so their candidate discovery states propagate together.
For a leaf $s$, compatibility is defined as $s \in S_i$.
Each internal node intersects the compatibility states of its children; if $m_i$ remains compatible, the node is added to the candidate list for $m_i$.
If no mutation remains compatible, the node cannot be reused by any update in the batch, and traversal does not continue to its parent.
The batched pass applies the same compatibility and pruning rules as independent discovery, while processing each active node once. Algorithm~\ref{alg:batched-candidate-discovery} summarizes the procedure.
Discovery does not modify topology; it reads the graph and writes only temporary per-node state and candidate lists.
For a child $c$ not reached by the active traversal, we define
$\operatorname{state}[c]=\varnothing$.
\begin{algorithm}[t]
\caption{Batched candidate discovery.}
\label{alg:batched-candidate-discovery}
\begin{algorithmic}[1]
\Procedure{DiscoverBatch}{$G,U$}
    \State Initialize $\operatorname{candidates}[i]$ for each mutation $m_i$
    \For{each sample leaf $s \in S_B$}
        \State $\operatorname{state}[s] \gets \{\,m_i \mid s \in S_i\,\}$
    \EndFor
    \State Process active nodes in reverse-topological order
    \For{each active internal node $n$}
        \State $\displaystyle \operatorname{state}[n] \gets \bigcap_{c \in \operatorname{children}(n)} \operatorname{state}[c]$
        \For{each $m_i \in \operatorname{state}[n]$}
            \State Add $n$ to $\operatorname{candidates}[i]$
        \EndFor
        \If{$\operatorname{state}[n] \neq \varnothing$}
            \State Activate the parents of $n$
        \EndIf
    \EndFor
    \State \Return $\operatorname{candidates}$
\EndProcedure
\end{algorithmic}
\end{algorithm}

\subsection{Bit-Parallel Mutation State}
\label{sec:bit-parallel-state}

\new{The recurrence of Eq.~\ref{eq:recurrence} is implemented using bit operations on a packed mask, with bit $i$ set exactly when $D(n) \subseteq S_i$.
For a leaf, the mask has one set bit per mutation the sample carries; for an internal node, the mask is the bitwise AND of its children’s masks, computing the intersection in a single word operation and evaluating the recurrence for many mutations simultaneously. 
For word width $w$, a batch of at most $w$ mutations needs one word of state per active node; larger batches use $q = \lceil k / w \rceil$ words, each covering a consecutive group of at most $w$ mutations and propagating independently.
Batch size is therefore not bounded by $w$: batches larger than $w$ cost $O(q)$ per node.}

\new{Because mutations in the same batch cannot reuse structure introduced by one another, deferred application may yield a less compact GRG than sequential mapping does.
In Table~\ref{tab:graph-compactness}, we quantify the trade-off.}

\subsection{Adaptive Carrier-Set Representation}
\label{sec:adaptive}

\new{Beyond the compatibility mask, the mapper maintains $D(n)$ for each candidate node, with $D(s) = \{s\}$ for a leaf and $D(n) = \bigcup_{c} D(c)$ for an internal node.
Because the GRG is a multitree with no diamond patterns, the descendant sets of distinct children are disjoint.
This disjointness is what makes accumulation cheap: sparse child vectors concatenate with no merge or duplicate checks, and dense children combine by bitwise OR with no overlaps to resolve.}

Carrier sets vary greatly in size: a rare variant may be carried by a handful of haplotypes, and a common one by most of the panel. 
Given that no single representation is efficient across this range, the approach adapts to each set’s density.
A small set is stored \emph{sparsely} as a vector of sample identifiers, which is efficient for the rare variants that dominate the workload. 
Once a set grows past a threshold $\tau$, it is stored
\emph{densely} as a width-$N$ bitset. 
Because propagation only adds samples, this conversion is one-way: dense sets combine with a bitwise OR, and when a sparse set is combined with a dense set, it is absorbed into the bitset.

\new{The threshold is defined by carrier-set density: a set with
$d = |D(n)|$ samples converts to dense form when $d/N \ge \tau$. The end-to-end experiments use $\tau = 1/32 = 0.03125$, the storage crossover for 32-bit identifiers, where a $d$-sample sparse vector ($32d$ bits) and a dense bitset ($N$ bits) cost the same:}
\begin{equation}
  32d \ge N \iff \frac{d}{N} \ge \frac{1}{32}.
\end{equation}
\new{The storage crossover need not minimize traversal time, since runtime also depends on memory traffic, cache behavior, and vectorization. 
In Section~\ref{sec:results}, we sweep $\tau$ to locate the runtime optimum.}

\new{The adaptive representation is complementary to batching.
Larger batches start from a bigger union of carrier samples and traverse a larger region of the graph, more often encountering descendant sets whose density exceeds $\tau$; dense bitsets therefore become increasingly valuable as batch size grows.
This sparse-or-dense choice is a classic technique in sparse linear algebra, \reviewed{mirroring a similar accumulator decision in Gustavson-style computation} ~\cite{gustavson1978, gilbert1992}, where a sparse accumulator stores only occupied entries and a dense accumulator becomes preferable once occupancy is high. 
This setting is a special case: the diamond-free GRG guarantees disjoint inputs and removes the duplicate detection that a general accumulator requires.}

\subsection{Mutation Updates}
\label{sec:applying-updates}

Discovery and application are kept separate.
Candidate discovery reads a fixed, read-only snapshot, and no graph changes occur until all candidate lists have been generated.
This separation is what lets a single shared traversal serve the entire batch: no mutation observes structure created by another in the same batch, so the shared pass needs no synchronization and never races on the topology. 
Once discovery completes, mutations are applied sequentially using the greedy attachment described in Section~\ref{sec:mutation-application}. 
Each mutation consumes its candidate list and either reuses a candidate that reaches exactly $S_i$ or builds new structure from compatible candidates and uncovered carriers, so the attachment node satisfies $D(n_i) = S_i$.
Because candidates are discovered before any are applied, a batch reuses preexisting structure but not any structure created within the batch; application proceeds in a fixed order, and the next batch discovers candidates on the updated graph, so later batches reuse nodes introduced earlier.

\begin{figure}
    \centering
    \begin{subfigure}[t]{0.48\linewidth}
        \centering
        \includegraphics[width=\linewidth]{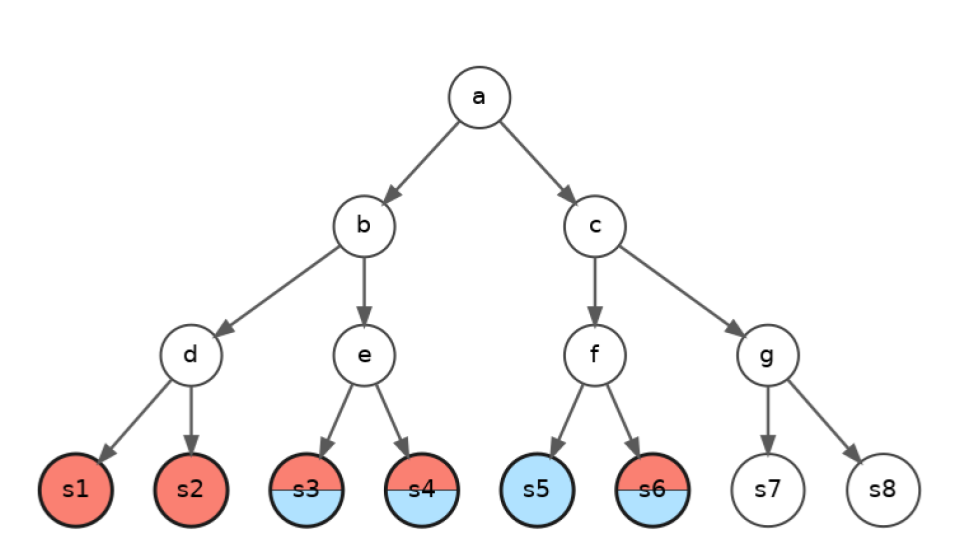}
        \caption{}
        \label{fig:multimm-p1}
    \end{subfigure}
    \hfill
    \begin{subfigure}[t]{0.48\linewidth}
        \centering
        \includegraphics[width=\linewidth]{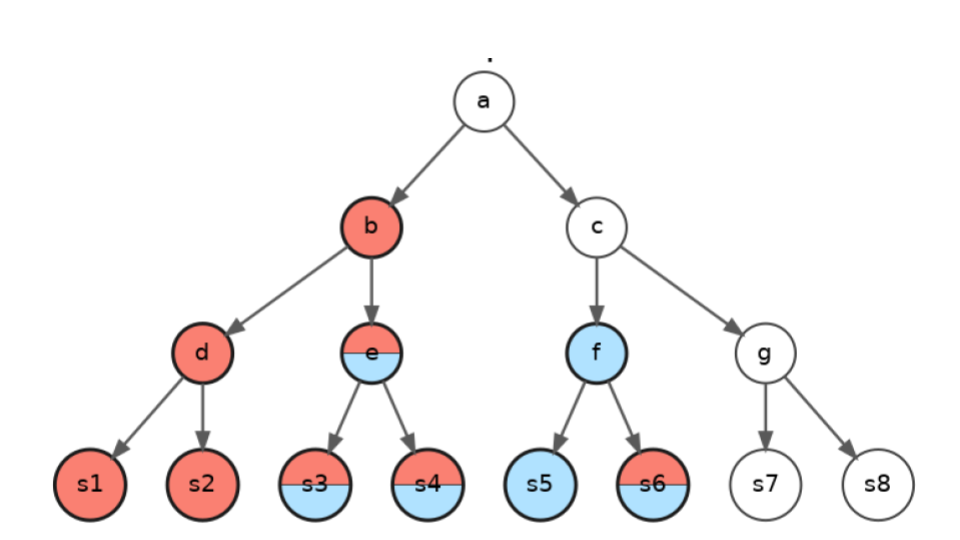}
        \caption{}
        \label{fig:multimm-p2}
    \end{subfigure}

    \vspace{0.5em}

    \begin{subfigure}[t]{0.48\linewidth}
        \centering
        \includegraphics[width=\linewidth]{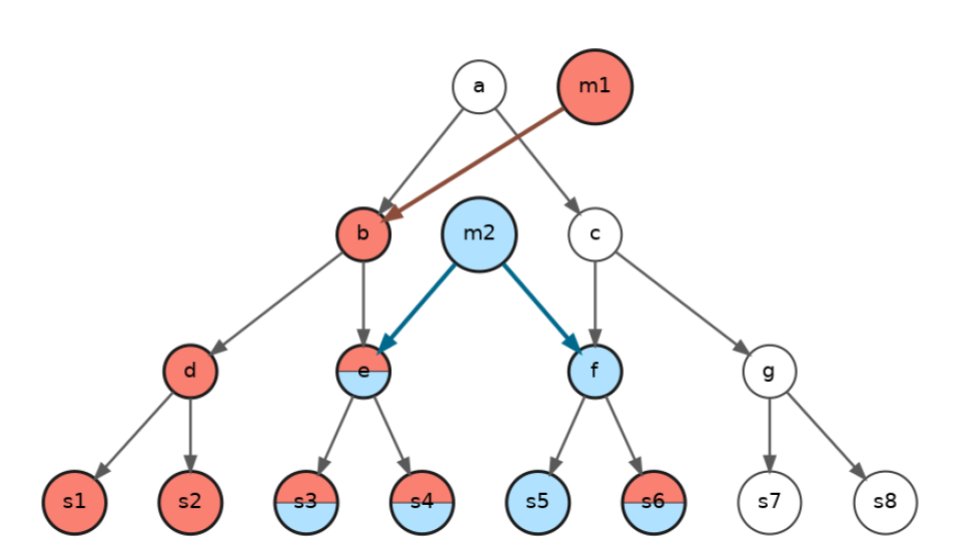}
        \caption{}
        \label{fig:multimm-p3}
    \end{subfigure}
    \hfill
    \begin{subfigure}[t]{0.48\linewidth}
        \centering
        \includegraphics[width=\linewidth]{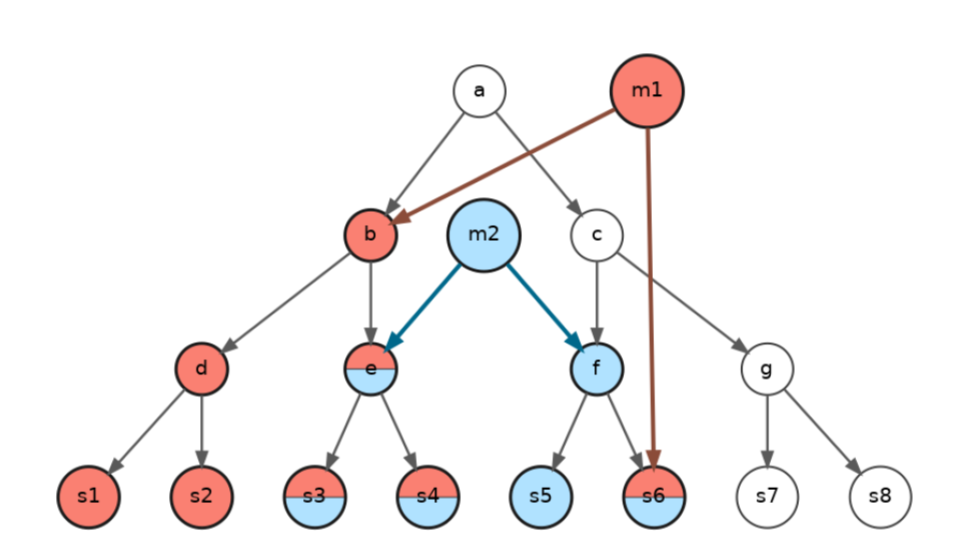}
        \caption{}
        \label{fig:multimm-p4}
    \end{subfigure}

    \caption{Batched mapping of two mutations with overlapping carrier sets.
    In \protect\subref{fig:multimm-p1}, the carrier sets of mutations $m_1$ and $m_2$ are shown in red and blue, respectively; split coloring indicates samples carrying both mutations. 
    In \protect\subref{fig:multimm-p2}, a single traversal over the union of the carrier sets propagates the state of both mutations.
    In \protect\subref{fig:multimm-p3}, the selected non-overlapping candidates are attached to $m_1$ and $m_2$. Finally, in \protect\subref{fig:multimm-p4}, any remaining uncovered carriers are attached directly, producing the exact carrier set for each mutation.}
    \label{fig:multimm}
\end{figure}

\subsection{Split-Based
Parallelism}\label{sec:split-base}

The existing implementation exposes coarse-grained parallelism by splitting the GRG by genomic position.
For each base-pair range, it builds a sub-GRG containing the mutations in that range, processes the sub-GRGs independently in parallel, and then merges the updated partitions.
\reviewed{Because each sub-GRG corresponds to a different genomic interval, updates to one sub-GRG do not modify mutations in another. Merging the independently processed sub-GRGs is thus straightforward: the independently updated partitions can be reassembled without having to resolve conflicting edits across partitions.}
Batching fits within this formulation, operating inside each concurrently processed sub-GRG, while splitting remains an outer layer.
It consolidates the mutation-specific traversals that would otherwise run independently on the same sub-GRG.

Each worker maintains its own sub-GRG along with the local state and modifications produced during mapping, and these modified sub-GRGs must be retained and merged, so adding workers increases peak memory.
Splitting adds workers at the cost of higher peak memory, whereas batching increases the work performed within each worker with little additional memory.
Batching is the memory-scalable complement to splitting, boosting throughput within each sub-GRG while using minimal additional memory.

\subsection{Correctness and Equivalence}
\label{sec:correctness}

Batched discovery produces, for every mutation, exactly the candidate list that independent discovery would produce on the same snapshot.
For a leaf $s$, $m_i$ is active exactly when $s \in S_i$; for an internal node $n$, $m_i$ remains active exactly when it is active at every child, which by Eq.~\ref{eq:recurrence} is equivalent to $D(n) \subseteq S_i$.
By induction over the reverse topological order, each candidate list is identical to its independent counterpart.
The application within a batch is serial and in a fixed order, so the output graph is deterministic; only its structure, not its carrier-set semantics, depends on that order.

\subsection{Complexity Analysis}
\label{sec:complexity-analysis}

\new{Let $E_i$ be the child edges examined while independently discovering candidates
for $m_i$, and let $Z_i$ be the \reviewed{number of} candidates generated \reviewed{for $m_i$}. 
Carrier-set propagation is excluded from the comparison, since it is identical under both schemes: both accumulate the same disjoint descendant sets along the same edges.
The independent discovery over $k$ mutations costs:
\begin{equation}
  O\!\left( \sum_{i=1}^{k} (|E_i| + Z_i) \right).
\end{equation}
Batched discovery, with $E_B = \bigcup_i E_i$ and $q = \lceil k/w \rceil$, costs:
\begin{equation}
  O(q\,|E_B| + Z_B), \qquad Z_B = \sum_{i=1}^{k} Z_i.
\end{equation}
The two differ only in how traversed child edges are counted: independent discovery examines an edge once per mutation whose traversal reaches it ($\sum_i |E_i|$), while batched discovery examines each distinct active edge once, at cost $q$ ($q\,|E_B|$).
Batching is beneficial when independent traversals overlap substantially.
The measured node-overlap factor $\rho(k)$ in Eq.~\ref{eq:overlap} provides an empirical proxy for this redundant traversal work.}

\new{The temporary mutation state requires $O(q A_B)$ words for $A_B$ simultaneously active nodes, plus descendant sets and candidate lists.
This scales with the active frontier of a single shared traversal, not with the number of concurrent workers.
That distinction is the source of batching’s memory scalability: splitting adds workers and replicates graph state, whereas batching increases per-worker work with a nearly fixed footprint
(Section~\ref{sec:split-base}).}

\subsection{Generating Polarization Updates}
\label{sec:generating-polarization-updates}

Post-construction updates in the evaluation come from allele polarization, generated per variant, reflecting how genotype data are refined in biobank pipelines~\cite{all2019all, bycroft2018uk}.
An ancestral reference is queried at each variant position; sites are skipped when the ancestral state is uncertain or the alleles are not single-nucleotide substitutions.
For a remaining site, if the ancestral allele matches the current reference, the site is already polarized and left unchanged; if it matches an alternate, the records at that position are rewritten relative to the new reference.
For a biallelic site, the replacement carrier set is the complement defined in Eq.~\ref{eq:complement}. 
If more than two alleles are present, alternate alleles are rewritten together.
The ancestral alternate becomes the new reference, the remaining alternates keep their carrier sets, and the former reference becomes a new alternate carried by the haplotypes that carry none of the original alternates.
Original records are removed, and the replacement from the update set is processed in batches.
\section{Results and Discussion}\label{sec:results}

\subsection{Experimental Setup}
\label{sec:experimental-setup}

Our batched mapping is evaluated on simulated GRGs generated with \texttt{stdpopsim} under the \texttt{OutOfAfrica\_2T12} demographic model for a European population \cite{adrion2020stdpopsim}.
\new{Datasets contain 5,000, 10,000,
20,000, 50,000, and 100,000 diploid individuals, corresponding to
10k, 20k, 40k, 100k, and 200k phased haplotypes, respectively (Table~\ref{tab:polarization-datasets}).}
The experiments were run on the NCSA Delta CPU partition. 
Each node has two AMD EPYC 7763 processors (128 physical cores) and 256 GB of DDR4-3200 RAM; each experiment used 1 node.

\begin{table}[t]
\centering
\caption{Dataset summary for polarization experiments.}
\label{tab:polarization-datasets}
\resizebox{\linewidth}{!}{
\begin{tabular}{lrrrr}
\toprule
Haplotypes & Nodes & Edges & Mutations & Remaps \\
\midrule
10k  & 4.2M  & 28.9M  & 1.3M & 134k \\
20k  & 6.7M  & 50.3M  & 1.9M & 188k \\
40k  & 10.4M & 86.6M  & 2.6M & 257k \\
100k & 17.5M & 181.3M & 3.7M & 372k \\
200k & 25.1M & 318.3M & 4.7M & 471k \\
\bottomrule
\end{tabular}
}
\end{table}

\begin{table}[t]
\centering
\caption{All of Us chromosome 21 GRG used for the real-data polarization experiment.}
\label{tab:aou-dataset}
\resizebox{\linewidth}{!}{
\begin{tabular}{lrrrr}
\toprule
Haplotypes & Nodes & Edges & Mutations &  Remaps \\
\midrule
829.7k & 37.2M & 487.1M & 12.0M & 85.8k \\
\bottomrule
\end{tabular}
}
\end{table}

\new{For each dataset, a GRG was generated along with its correct ancestral reference.
Given that polarizing against that reference would be a no-op, a modified reference was produced by flipping the ancestral allele at 10\% of sites corresponding to biallelic mutations in the GRG. \reviewed{The 10\% flip rate was chosen to generate a large, controlled remapping workload, enabling consistent performance comparisons across simulated datasets.} 
Polarizing against the modified reference induces a controlled remapping workload exactly at those sites, and because the flipped sites are known by construction, the expected complements are known in advance, allowing verification of the resulting GRG.}

\new{Polarization was also evaluated on human chromosome 21 from the All of Us (AoU) Research Program~\cite{all2019all}, an NIH cohort study of a diverse US population, using a chromosome-specific GRG (Table~\ref{tab:aou-dataset}) polarized against the Ensembl ancestral reference genome~\cite{paten2008genome}.}

\edit{For the AoU experiment, chromosome 21 GRG was divided into 194 sub-GRGs using a 200 kbp window.
The experiment ran in the AoU Researcher Workbench, the only platform where this controlled-access dataset can be used, on a Google Cloud \texttt{n2-highmem-32} instance with an Intel Cascade Lake CPU, 16 cores, 32 hardware threads, and 256 GB of memory.
Other parameters matched the simulated experiments, except up to 4 concurrent single-threaded jobs instead of 8.}

\new{Polarization ran with batch sizes 1, 16, 64, 256, and 1024. 
As described in Section~\ref{sec:bit-parallel-state}, this spans both the single-word and multi-word regimes; with 64-bit words, a batch size of 1024 requires $q = 16$ words of state per node.
Other parameters were held fixed. Using the split-based procedure of Section~\ref{sec:split-base}, each GRG was divided into 200-kbp sub-GRGs, and up to eight \reviewed{single-threaded workers processed sub-GRGs concurrently within each experiment, with thread placement managed by Slurm. Different experiments were run separately}. The $\approx 51$-Mbp simulated region from chromosome 22 produced 255 sub-GRGs per dataset.}

\new{For each configuration, end-to-end wall-clock time and peak memory usage were recorded, along with split and per-partition processing times.
In the mapping, graph traversal and application times were measured separately, along with per-mutation node visits and distinct nodes visited per batch. 
A dense-membership $\tau$ sweep evaluated when dense bit-vector membership beats sparse membership during traversal. To keep it tractable, the sweep ran on the first 200-kbp split from each dataset with $\tau \in \{0.001, 0.005, 0.01, 0.03125, 0.05, 1.00\}$, where $\tau = 1.00$ disables dense membership as the sparse baseline.
For each dataset–batch pair, the best non-baseline $\tau$ is reported as a traversal time speedup over that baseline.}

\subsection{Polarization Performance}
\label{sec:polarization-performance}

\begin{figure*}[t]
    \centering
    \includegraphics[width=\linewidth]{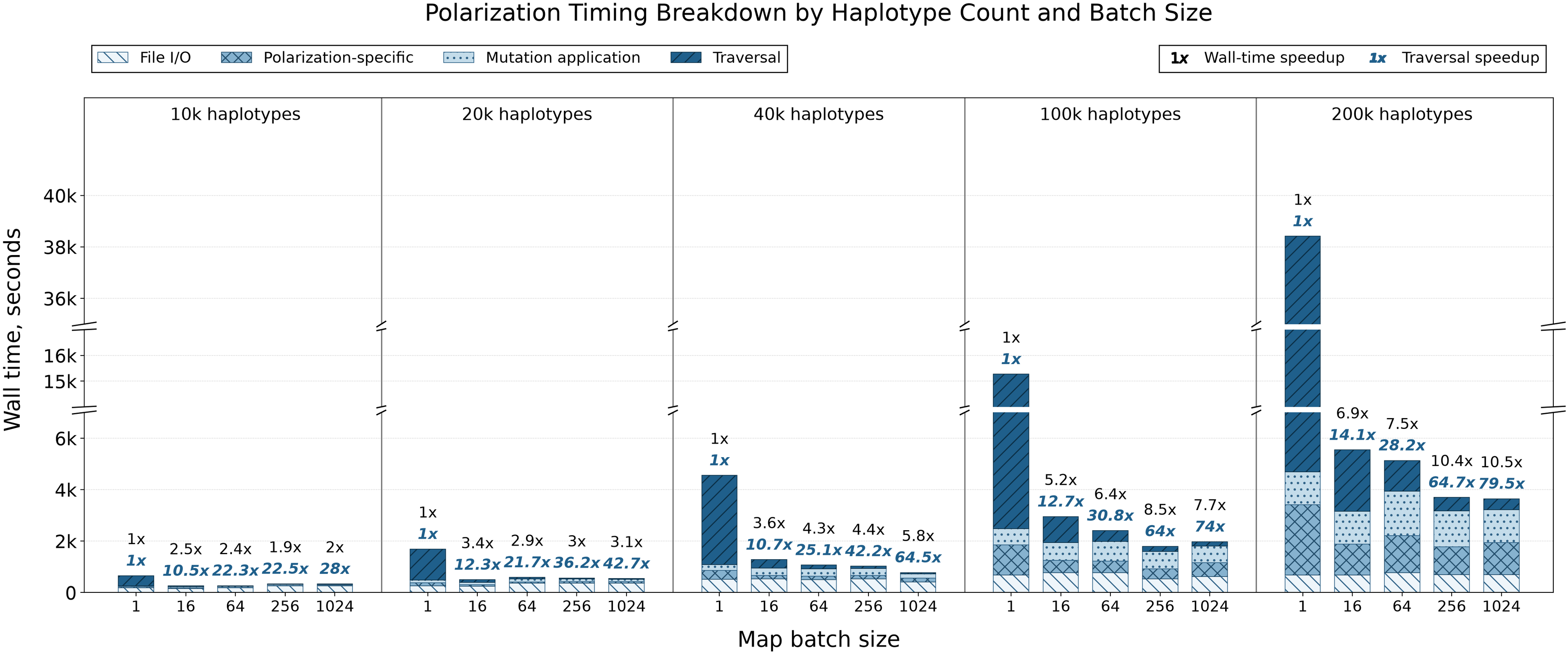}
    \caption{Polarization runtime breakdown across haplotype counts and batch sizes. Larger batches reduce traversal time, while mutation application, polarization-specific processing, and file I/O increasingly limit end-to-end speed.}
    \label{fig:timings}
\end{figure*}

\new{Figure~\ref{fig:timings} shows that batching consistently reduces the end-to-end polarization time.
The best observed speedup increases with dataset size: $2.5\times$ at 10k haplotypes, increasing to $3.4\times$, $5.8\times$, and $8.5\times$, and reaching $10.5\times$ at 200k haplotypes. \reviewed{On smaller datasets, pipeline overhead and run-to-run variability dominate end-to-end time.
As a result, wall time grows with batch size in some settings even though traversal gets faster.}
Larger datasets amortize repeated traversal more, and therefore benefit more.
Traversal time dominates at batch size 1 but falls sharply with batching; at batch size 1024, traversal time drops by $28.0\times$ at 10k and $79.5\times$ at 200k.
Mutation application, polarization-specific work, and I/O then account for a growing fraction of the runtime, so the end-to-end speedup stays below the traversal speedup, and additional batching yields diminishing returns.} 

\begin{figure}
    \centering
    \includegraphics[width=1\linewidth]{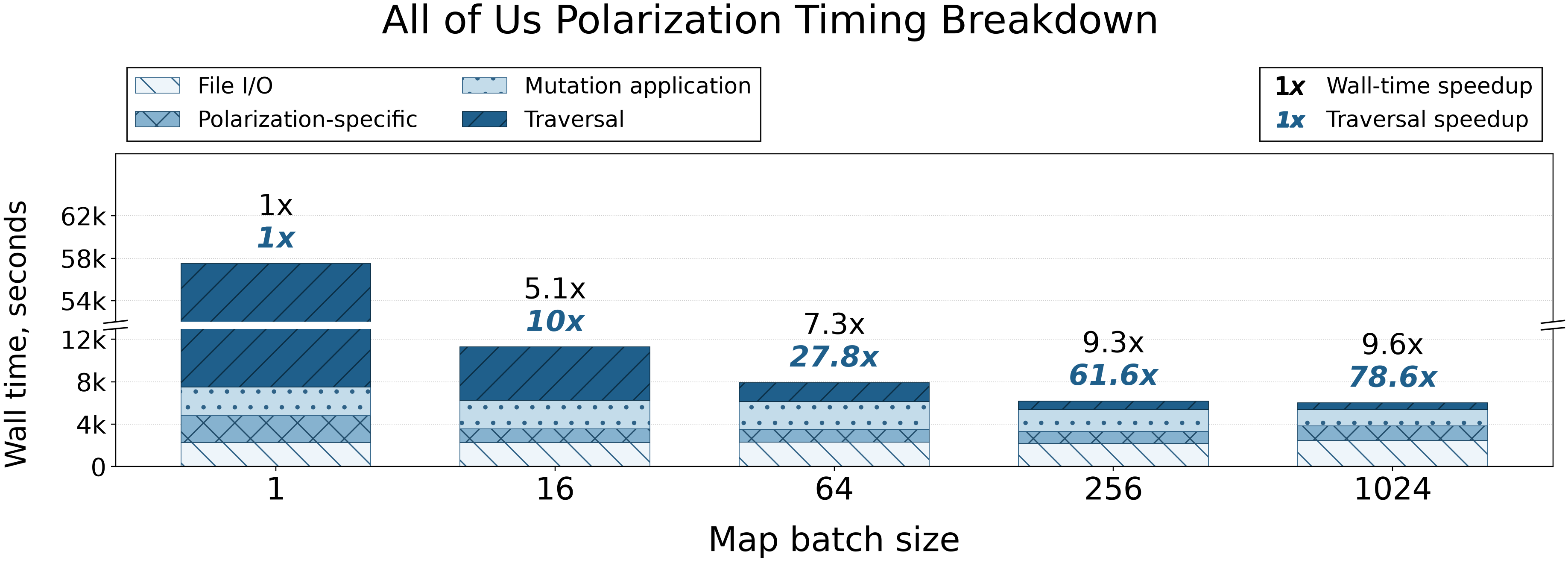}
    \caption{Polarization runtime breakdown across batch sizes on the All of Us dataset~\cite{all2019all}. 
    The speedups and timing distributions are similar to those for the large 200k haplotype dataset.}
    \label{fig:placeholder}
\end{figure}

The All of Us experiment shows the same trend on real data: batching substantially reduces traversal time and yields a $9.6\times$ end-to-end speedup at a batch size of 1024, with nontraversal phases accounting for most of the remaining runtime.

In the simulated data, the smaller datasets reach diminishing returns at modest batch sizes, whereas the 40k and 200k datasets continue to improve up to a batch size of 1024. The 100k dataset reaches its lowest wall-clock time at a batch size of 256, though the gap to 1024 is small and likely due to run-to-run variation.

\paragraph{Traversal Reuse}

To quantify repeated traversal, let $V_i$ be the nodes visited when $m_i$ is mapped independently. 
For a batch of $k$ mutations, the overlap factor is:
\begin{equation}
  \rho(k) = \frac{\sum_{i=1}^{k} |V_i|}{\left| \bigcup_{i=1}^{k} V_i \right|}.
  \label{eq:overlap}
\end{equation}
This is the average number of independent visits per distinct node:
$\rho(k) = 1$ means no overlap, and larger values mean more repeated work is available to eliminate.
Results are reported on the 200k dataset, the largest one evaluated and the one showing the greatest reuse.
The overlap factor rises from $1.0\times$ at batch 1 to $15.8\times$, $62.4\times$, $232.4\times$, and $916.8\times$ as batch size grows, so at batch 1024 independent mapping would process each distinct node about 917 times on average.
The nearly proportional increase confirms that mutations within a batch traverse largely shared regions. 
The measured traversal speedup stays below the overlap factor because the batched pass must propagate mutation-specific state and perform bookkeeping for each mutation.

\begin{figure}[t]
    \centering
    \includegraphics[width=\linewidth]{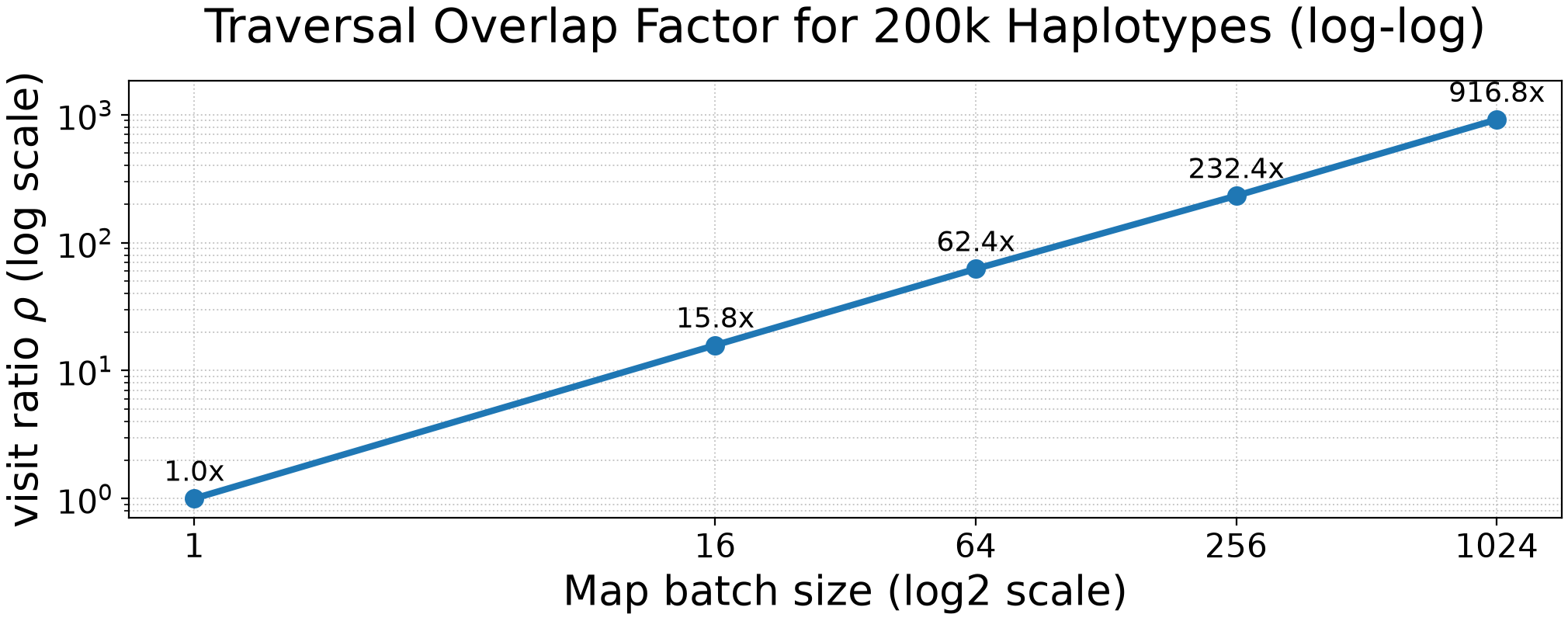}
    \caption{The overlap factor \(\rho\) (computed as \(\frac{\text{independent visits}}{\text{shared visits}}\)) for the 200k haplotype dataset. 
    Overlap grows nearly proportionally with batch size, reaching \(916.8\times\) at a batch size of 1024.}
    \label{fig:overlap-factor}
\end{figure}

\paragraph{Memory Usage}

\begin{table}[t]
\centering
\caption{Peak memory by haplotype count and batch size. Batch 1 is reported in GiB; the remaining columns report peak memory relative to Batch 1 for the same dataset. Even at batch size 1024, peak memory stays within \(1.29\times\) of the baseline.}
\label{tab:memory-by-batch}
\resizebox{\linewidth}{!}{%
\begin{tabular}{rrrrrr}
\toprule
Total haplotypes
    & Batch 1 (GiB)
    & Batch 16
    & Batch 64
    & Batch 256
    & Batch 1024 \\
\midrule
10k  & 4.40   & $1.051\times$ & $0.987\times$ & $0.986\times$ & $1.109\times$ \\
20k  & 7.48   & $1.035\times$ & $1.047\times$ & $1.199\times$ & $1.252\times$ \\
40k  & 16.40  & $0.958\times$ & $0.986\times$ & $1.110\times$ & $1.279\times$ \\
100k & 51.92  & $1.003\times$ & $1.028\times$ & $1.105\times$ & $1.190\times$ \\
200k & 129.05 & $0.999\times$ & $1.004\times$ & $1.064\times$ & $1.286\times$ \\
\bottomrule
\end{tabular}%
}
\end{table}

The gains require only a moderate increase in memory. Table~\ref{tab:memory-by-batch} reports batch-1 peak memory in GiB and peak memory for larger batches relative to that baseline.
Batches 16 and 64 stay close to the batch-1 footprint. 

\reviewed{Peak memory rises slightly at larger batch sizes because the shared traversal maintains state for more mutations and typically spans a larger union of carrier samples. This increases both the per-node mutation masks and the temporary memory used to propagate and merge descendant sample sets during traversal.} However, even the largest batch, batch 1024, uses at most $1.286\times$ baseline.
On the 200k dataset, that is an increase from 129.05 GiB to roughly 166 GiB while delivering a $10.5\times$ end-to-end speedup.
The values slightly below $1.0\times$ reflect variation in process-level peak memory measurement rather than a systematic reduction.

\paragraph{Adaptive Carrier Set Membership}

\begin{figure}[t]
    \centering
    \includegraphics[width=\linewidth]{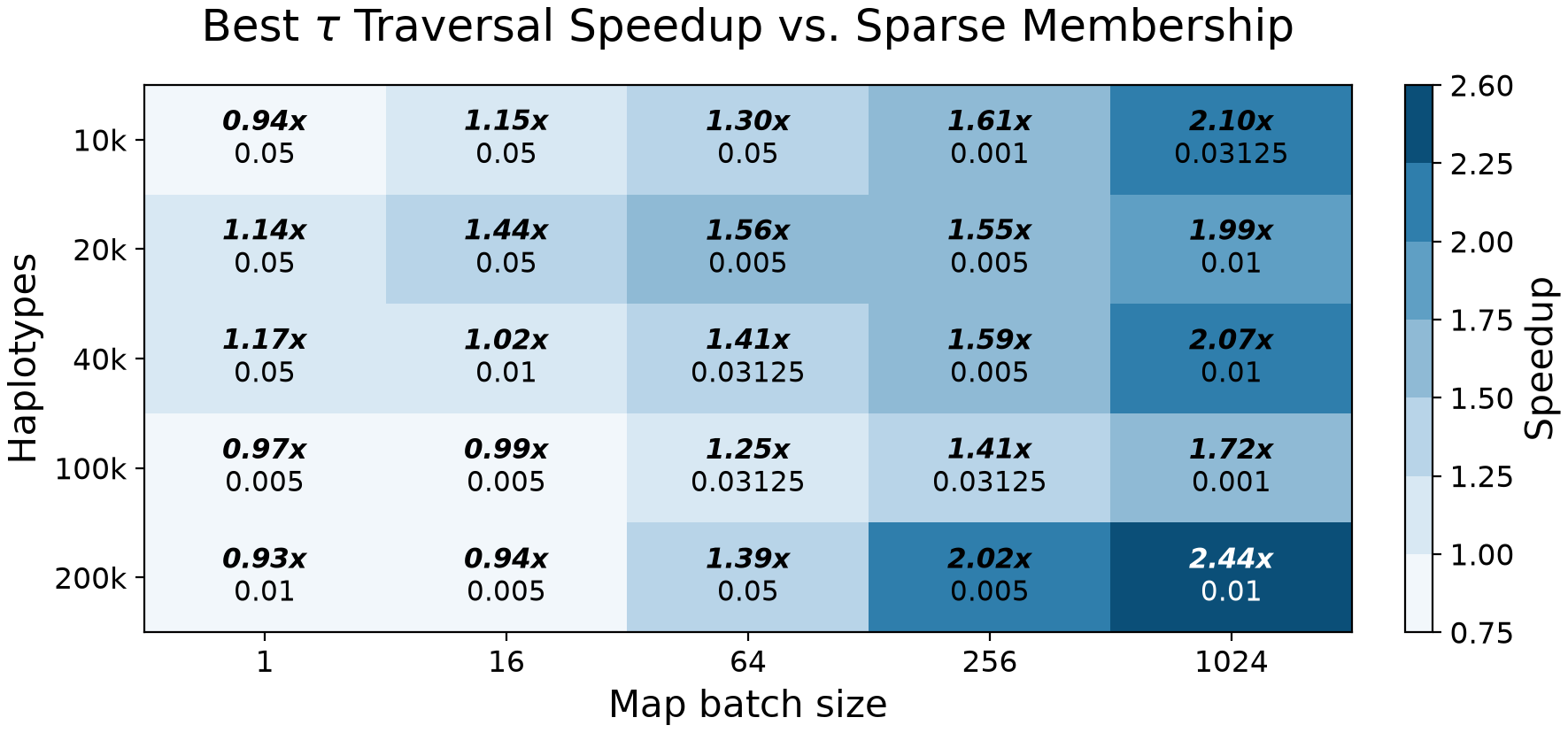}
    \caption{The best traversal speedup from adaptive sparse/dense membership across datasets and batch sizes. The sweep was performed on the first 200-kbp split only. Each cell reports the speedup over sparse-only membership and the value of \(\tau\) with the lowest traversal time.
    Dense membership provides the greatest benefits at larger batch sizes; the optimal \(\tau\) varies.}
    \label{fig:membership-cutoff}
\end{figure}

\new{Figure~\ref{fig:membership-cutoff} reports the best traversal speedup from the dense-membership $\tau$ sweep, with each cell comparing the fastest non-baseline $\tau$ to sparse-only membership. 
Dense membership helps little at small batch sizes and can even be slower because of conversion overhead, but its benefit grows with batch size: at batch size 1024, every evaluated $\tau$ outperforms sparse-only on every dataset, and selecting the best $\tau$ yields a $1.72\times$–$2.44\times$ traversal speedup across datasets.
This reinforces that batching and adaptive representation complement each other, because larger batches trigger broader traversals and encounter dense descendant sets more often, creating more opportunities for bit-parallel dense accumulation to outperform sparse vectors.
The runtime-optimal $\tau$ varies, but a fixed $\tau = 0.01$ stays within 20\% of the tuned optimum for 22 of 25 configurations and is only 9.5\% slower on average, so most of the benefit is available without per-workload tuning.}

\paragraph{Output Graph Compactness}
\label{sec:graph-compactness}

Because structural updates are deferred until discovery is complete for a batch, a mutation cannot reuse structure introduced by another mutation in the same batch. 
This does not change the carrier sets represented, but it may produce a less compact GRG.
In addition, although the existing mapper can compact the graph by locally pattern-matching newly created structures against existing nodes, batching can introduce similar structures across multiple newly created nodes that are not immediately visible to this localized procedure.
The effect is quantified on the 200k dataset, which consistently shows the largest edge-count and serialized-size overhead (Table~\ref{tab:graph-compactness}); all smaller datasets show equal or lower overhead. 
The node count is effectively unchanged across batch sizes.
A substantial loss of compactness occurs primarily at batch size 1024, where the edge count increases by 78.24\% and the serialized size by 27.15\%.
Deferred updates have little effect at small and moderate batch sizes, while the most aggressive configuration trades some compactness for higher mapping performance. Future work could analyze structural similarity among nodes created within a batch to identify additional opportunities for sharing and further reduce compactness loss.

\begin{table}[t]
\centering
\caption{Output graph structure for the 200k-haplotype dataset. Batch 1 is reported using absolute values; the remaining rows report changes relative to Batch 1.}
\label{tab:graph-compactness}
\resizebox{0.9\linewidth}{!}{%
\begin{tabular}{rrrr}
\toprule
Batch size & Nodes & Edges & Serialized size \\
\midrule
1    & 25.1M          & 318.3M          & 1.057 GB \\
16   & $+0.00\%$      & $+1.00\%$       & $+0.38\%$ \\
64   & $+0.02\%$      & $+4.34\%$       & $+1.51\%$ \\
256  & $+0.03\%$      & $+18.73\%$      & $+6.43\%$ \\
1024 & $+0.01\%$      & $+78.24\%$      & $+27.15\%$ \\
\bottomrule
\end{tabular}%
    }
\end{table}
\section{Related Work}

Genetic variation is commonly stored in compressed dense formats such as VCF~\cite{danecek2011variant_vcf} and BCF~\cite{li2011statistical_bcf}, which do not scale well with sample size.
PLINK PGEN improves on this by storing data sparsely and compressing shared structure in place~\cite{plink}, but such formats still treat this sharing as flat and ignore its underlying genealogical hierarchy.
The recently introduced genotype representation graph~\cite{dehaas2024grg, dehaas2026general} instead shares subgraph structure across samples, placing mutations on that shared structure for lossless compression.
Mikado~\cite{mikado} formulates computation on this structure using sparse linear algebra.
Under a reverse topological ordering, the GRG adjacency matrix is strictly block-lower-triangular, so genotype matrix–vector products become blocked sparse triangular solves, accelerating computation on GRG by orders of magnitude.
These analyses treat the graph as fixed; the present work addresses the complementary problem of editing it after it has been constructed.

Running many searches on a single graph is the setting for multi-source breadth-first search (BFS) traversal.
MS-BFS~\cite{then2014more} and iBFS~\cite{liu2016ibfs} jointly execute concurrent breadth-first searches, combining their frontiers and attaching compact per-search state to each visited vertex so that a shared vertex is expanded once for every search that reaches it.
These methods target concurrent BFS, in which searches reach a vertex at different levels, must maintain per-search discovery state, and repeatedly rebuild level-aligned frontiers.
In contrast, our remapping uses different traversal semantics.
Each search starts from a carrier set and proceeds upward through an acyclic graph, so per-mutation visit sets can be combined in a single reverse topological pass without level alignment.
Moreover, in concurrent BFS, a shared frontier reduces memory traffic but not total work, because each search still inspects each vertex independently. 
In our batched discovery process, a shared compatible node is processed only once for the entire batch, reducing the total work.

Propagating and merging carrier sets up the DAG is a sequence of set union operations whose cost depends on operand density.
This is similar to the sparse-versus-dense accumulator choice at the core of Gustavson-style sparse matrix computation~\cite{gustavson1978}.
A sparse accumulator stores only occupied entries, whereas a dense accumulator uses a full-width representation, which is preferable once occupancy is high~\cite{gilbert1992}.
Recent SpGEMM implementations dynamically choose the accumulator based on density~\cite{ocean, parger2020speck}.
Our adaptive carrier-set representation makes an analogous choice, simplified by the diamond-free GRG structure, which guarantees disjoint inputs and removes the need for duplicate detection during sparse accumulation.
This situates GRG editing within a broader body of work at the intersection of sparse linear algebra and biology~\cite{bella, elba, pastis, pastisplus, jaccardgenome, hipmcl}.
\section{Conclusion}
\label{sec:conclusion}

\new{This paper presented a batched mutation-mapping algorithm for GRGs that replaces independent reverse-topological traversals with a single traversal.
Bit-parallel operations maintain compact mutation-specific state, adaptive selection switches between sparse and dense carrier set memberships, and the method integrates with the existing split-based parallel pipeline.
Its effectiveness comes from exploiting the substantial overlap between independent traversals.
These irregular, pointer-chasing passes have data-dependent frontiers whose shape varies with carrier-set density, and batching consolidates them into a single shared traversal that eliminates repeated work without modifying the graph concurrently.}

In addition, we applied the proposed algorithm to allele polarization, where batching becomes increasingly beneficial at larger scales.
On the largest dataset (200k haplotypes, 4.7M mutations), it reduces shared traversal time by $79.5\times$ and end-to-end polarization time by $10.5\times$, while keeping peak memory within $1.3\times$ of the batch-1 baseline.
The adaptive dense membership adds an additional $1.7\times$–$2.4\times$ traversal improvement at large batch sizes, with its benefit increasing as batch size grows and larger shared traversals encounter dense descendant sets more frequently.
Deferring structural updates has little effect on compactness for small and moderate batch sizes, although the most aggressive configuration trades some sharing for improved mapping performance.

As shared traversal becomes faster, mutation application, polarization-specific work, and I/O dominate end-to-end performance, motivating future work on these stages.
Nevertheless, our results demonstrate the effectiveness of collective traversal for a high-overlap, multi-query workload on GRGs and suggest broader applicability to graph searches with composable per-task state.

\section*{Acknowledgment}
The authors thank the members of the ALPS Lab at Cornell University for their feedback and discussion. 
This research used resources from the National Energy Research Scientific Computing Center, a DOE Office of Science User Facility supported by the Office of Science of the U.S. Department of Energy under Contract No. DE-AC02-05CH11231, using NERSC award ASCR-ERCAP0030076. This material is based upon work supported by the National Science Foundation under Grant IIS-2435801. 
This work used the DeltaAI system at the National Center for Supercomputing Applications (NCSA) through allocation CIS251351 from the Advanced Cyberinfrastructure Coordination Ecosystem: Services \& Support (ACCESS) program, which is supported by U.S. National Science Foundation grants \#2138259, \#2138286, \#2138307, \#2137603, and \#2138296.
The authors gratefully acknowledge All of Us participants for their contributions, without whom this research would not have been possible.
In addition, we thank the National Institutes of Health All of Us Research Program for making available the participant data examined in this study. 
This study used data from the All of Us Research Program Controlled Tier Dataset CDRv8, available to authorized users on the Researcher Workbench.

\bibliographystyle{IEEEtran}
\bibliography{ref}
\clearpage
\appendix
\section{Artifact Description}

\subsection{Artifact Identification}
The artifact supports the paper’s evaluation of batched mutation remapping in genotype representation graphs (GRGs).
GRGL implements the mapping algorithm, and GRAPP provides the allele-polarization workflow.
The \href{https://github.com/CornellHPC/GRG-shared-traversal}{GRG-shared-traversal repository} includes GRGL and GRAPP as pinned Git submodules, along with a generator for simulated datasets.

\subsection{Hardware}

The simulated experiments ran on the platform described in Section IV-A, using eight concurrent polarization jobs and one mapping thread per job.

\subsection{Input Data}

The simulated experiments use chromosome 22 GRGs with 10k, 20k, 40k, 100k, and 200k haplotypes, each paired with an ancestral FASTA and a modified FASTA.
The included \texttt{generate\_data.py} script produces these inputs.
To construct GRGs from other sources, refer to the GRGL documentation.
The chromosome 21 experiment uses real human genetic data from the All of Us Research Program. 
Because its input GRG is controlled access, it is not part of the public artifact.
The authorized users conducted this experiment within the All of Us Researcher Workbench.

\subsection{Output}

Input generation produces a tree sequence, a GRG, two FASTA files, a list of flipped sites, and a summary of the generation parameters and output paths.
Polarization produces an output GRG, a CSV of per-batch mapping timings and node visits, and per-phase timings written to standard error.

\section{Artifact Evaluation}

\subsection{Installation}

Clone the artifact together with its submodules.
\begin{quote}
\small\ttfamily
git clone --recursive\\
\url{https://github.com/CornellHPC/GRG-shared-traversal.git}
\end{quote}
Then follow the repository README to install GRGL and GRAPP, in
that order, into the same Python environment. 
Input generation
additionally requires \texttt{stdpopsim} and \texttt{tskit}.

\subsection{Generating Simulated Inputs}

The generator uses the \texttt{EUR} population from the human \texttt{OutOfAfrica\_2T12} model.
The following command generates the 10k-haplotype dataset for the full chromosome~22.
\begin{quote}
\small\ttfamily
python generate\_data.py --out data/10k\\
\hspace*{1em}--samples 5000 --chromosome chr22\\
\hspace*{1em}--flip-pct 0.1 --seed 1
\end{quote}
Because \texttt{--samples} counts diploid individuals, the 20k, 40k, 100k, and 200k datasets use values of 10,000, 20,000, 50,000, and 100,000, respectively. 
The script runs \texttt{stdpopsim}, converts the resulting \texttt{.trees} file to a GRG, and writes four files: \texttt{polarized.grg} (the input GRG for polarization), \texttt{true.fa} (the original reference FASTA), \texttt{fake.fa} (the FASTA with flipped sites), and \texttt{flips.tsv} (the list of flipped sites).
It also records the generation parameters and output paths in \texttt{summary.tsv}.

\subsection{Running Polarization}

The experiments evaluate mapping batch sizes $k \in \{1, 16, 64, 256, 1024\}$ while holding all other parameters fixed. 
The command below runs the $k = 256$ configuration.
Create the \texttt{results/tmp} directory in advance, and use a separate results directory for each configuration to avoid overwriting outputs.
\begin{quote}
\small\ttfamily
grapp polarize data/10k/polarized.grg\\
\hspace*{1em}data/10k/fake.fa\\
\hspace*{1em}-o results/output.grg\\
\hspace*{1em}--map-batch-size 256\\
\hspace*{1em}--map-input-batch-size 4096\\
\hspace*{1em}--thread-count 1\\
\hspace*{1em}--dense-membership-cutoff 0.03125\\
\hspace*{1em}--jobs 8\\
\hspace*{1em}--split-threshold 200000\\
\hspace*{1em}--keep-no-match\\
\hspace*{1em}--temp-dir results/tmp\\
\hspace*{1em}--map-timing-csv results/timing.csv\\
\hspace*{1em}> results/polarize.log\\
\hspace*{1em}2> results/polarize.err
\end{quote}
The \texttt{--keep-no-match} option preserves mutations at positions marked as unknown in the generated FASTA files.
The \texttt{--map-input-batch-size} option groups mutation updates before passing them to the mapper, which amortizes per-call overhead such as file I/O and communication between Python and C++.
This input batching is independent of the mapping batch size $k$ used in the experiments.

\subsection{Reproducing and Analyzing Results}

To verify correctness, export the input and output GRGs to VCF.
\begin{quote}
\small\ttfamily
grapp export data/10k/polarized.grg\\
\hspace*{1em}--vcf results/input.vcf --contig chr22\\
grapp export results/output.grg\\
\hspace*{1em}--vcf results/output.vcf --contig chr22
\end{quote}
For every position listed in \texttt{flips.tsv}, swap the reference and alternate alleles, while keeping the allele carried by each sample unchanged.
Leave all other sites unchanged.

For performance analysis, the mapping CSV reports traversal time, mutation application time, and node visits for each batch. 
The standard error log reports GRG loading and saving times, as well as the times for splitting, processing parts, and merging. 
Because these measurements include nested phases and concurrently processed parts, they do not sum to the total runtime and are best used to understand how processing time is distributed. 
To measure total wall-clock time and peak memory usage,
run the command with an external tool such as
\texttt{/usr/bin/time}.
\end{document}